\documentclass[11pt]{article}
\usepackage{moriond,epsfig}
\def\al{\alpha}

\def\be{\begin{equation}}
\def\ee{\end{equation}}
\def\bea{\begin{eqnarray}}
\def\eea{\end{eqnarray}}

\begin{document}
\vspace*{4cm}
\title{Recent electroweak measurements from the {H1} and {ZEUS} experiments}

\author{ Arafat Gabareen Mokhtar}
\address{Tel Aviv University,\\
  Tel Aviv, Israel \\
  for the {H1} and {ZEUS} collaborations\\
  Email: gabareen@mail.desy.de}

\maketitle\abstracts{ Neutral-current ({NC}) and charged-current
  ({CC}) deep inelastic scattering ({DIS}) reactions have been studied
  in $e^+p$ and $e^-p$ collisions using the {H1} and {ZEUS} detectors
  at {HERA I}. Following the upgrade of the {HERA} accelerator, the
  {HERA II} program recently started with the first data in $e^+p$
  scattering with longitudinally polarised positrons. In this paper,
  a summary of the electroweak results from {HERA I} and the first
  measurement of the cross section for $e^+p$ CC {DIS} at a
  longitudinal polarisation value of $33\%$ from the {ZEUS}
  collaboration are presented.}

\section{HERA I}

\label{subsec:kine}

At {HERA}~\cite{HERA}, a proton beam of
$920~\mathrm{GeV}$~\footnote{Until 1998, the energy of the proton
  beams were $820~\mathrm{GeV}$. } collides with an electron or
positron~\footnote{In the following, for simplicity we will denote as
  electron the charged lepton, independent of whether it is an $e^+$
  or an $e^-$, unless otherwise stated.} beam of $27.5~\mathrm{GeV}$.
The electron-proton ($ep$) interactions proceed via $\gamma$ and
$Z^0$- exchange in the neutral current ({NC}) reaction, or via
$W^{\pm}$ exchange in the charged current ({CC}) interaction. The
description of events is usually given in terms of the $ep$ center of
mass energy squared, $s$, and two out of three Lorentz invariant
quantities, $Q^2$: the absolute value of the invariant mass squared of
the exchanged particle, $x$: the fraction of the proton momentum
carried by the struck quark, and $y$; the fractional energy
transferred to the proton in its rest frame.  These variables are
related through $Q^2=sxy$, if the masses of the electron and the
proton are neglected.

\subsection{Neutral current}

\label{subsec:prod}

The double differential {NC} cross section for the reaction
$ep\rightarrow eX$ with longitudinally unpolarised electron is given
by

\begin{eqnarray}
        \frac{d^2\sigma_{Born}^{NC}(e^{\pm}p)}{dxdQ^2} 
        = \frac{2\pi\al^2}{xQ^4}
        \left [Y_+F_2^{NC}(x,Q^2)\mp Y_-xF_3^{NC}(x,Q^2)\right ]\, ,
\label{eq:dsnc}
\end{eqnarray}
where $\alpha$ denotes the fine-structure constant and $Y_{\pm}=1\pm
(1-y)^2$. The contribution of the longitudinal structure function was
neglected. The structure function $F_2^{NC}$ can be divided into three
terms, due to electromagnetic interactions, to the interference
between the photon and $Z^0$ exchange, and pure $Z^0$
exchange~\cite{Klein}, such that

\begin{eqnarray}
        F_2^{NC} \equiv {\bf F_2^{em}} 
        -v_e\frac{\kappa Q^2}{Q^2+M_{Z^0}^2}{\bf F_2^{\gamma {Z^0}}}
        + (v_e^2+a_e^2)\left (\frac{\kappa Q^2}{Q^2+M_{Z^0}^2}\right )^2{\bf F_2^{Z^0}}\, ,  
\end{eqnarray}
where $\kappa \equiv
4\frac{M_W^2}{M_{Z^0}^2}(1-\frac{M_W^2}{M_{Z^0}^2})$, $M_W$
($M_Z^0$) is the $W^{\pm}$ ($Z^0$) boson mass, and $v_e$ and $a_e$ are
the {NC} couplings of the $Z^0$ to the electron.  The couplings of the
quarks are contained in the respective $F_2^{NC}$ structure functions
($F_2^{em}$,$F_2^{\gamma Z^0}$ and $F_2^{Z^0}$).

The parity violating structure function, $xF_3^{NC}$, contains two
terms: one due to the interference between photon and $Z^0$ exchange
amplitudes and a term due to pure $Z^0$ exchange,
\begin{eqnarray}
        xF_3^{NC} \equiv 
        -a_e\frac{\kappa Q^2}{Q^2+M_{Z^0}^2}{\bf xF_3^{\gamma Z^0}} 
        + 2v_ea_e\left (\frac{\kappa Q^2}{Q^2+M_{Z^0}^2}\right )^2{\bf xF_3^{Z^0}} \, .
\end{eqnarray}

When $\frac{Q^2}{M_{Z^0}^2}\ll 1$, the {NC} process is dominated by
photon exchange. When $Q^2$ becomes of the order of $M_{Z^0}^2$, the
contribution from the interference and pure $Z^0$ exchange is
non-negligible.

\begin{figure}
\begin{minipage}{8.0cm}
\psfig{figure=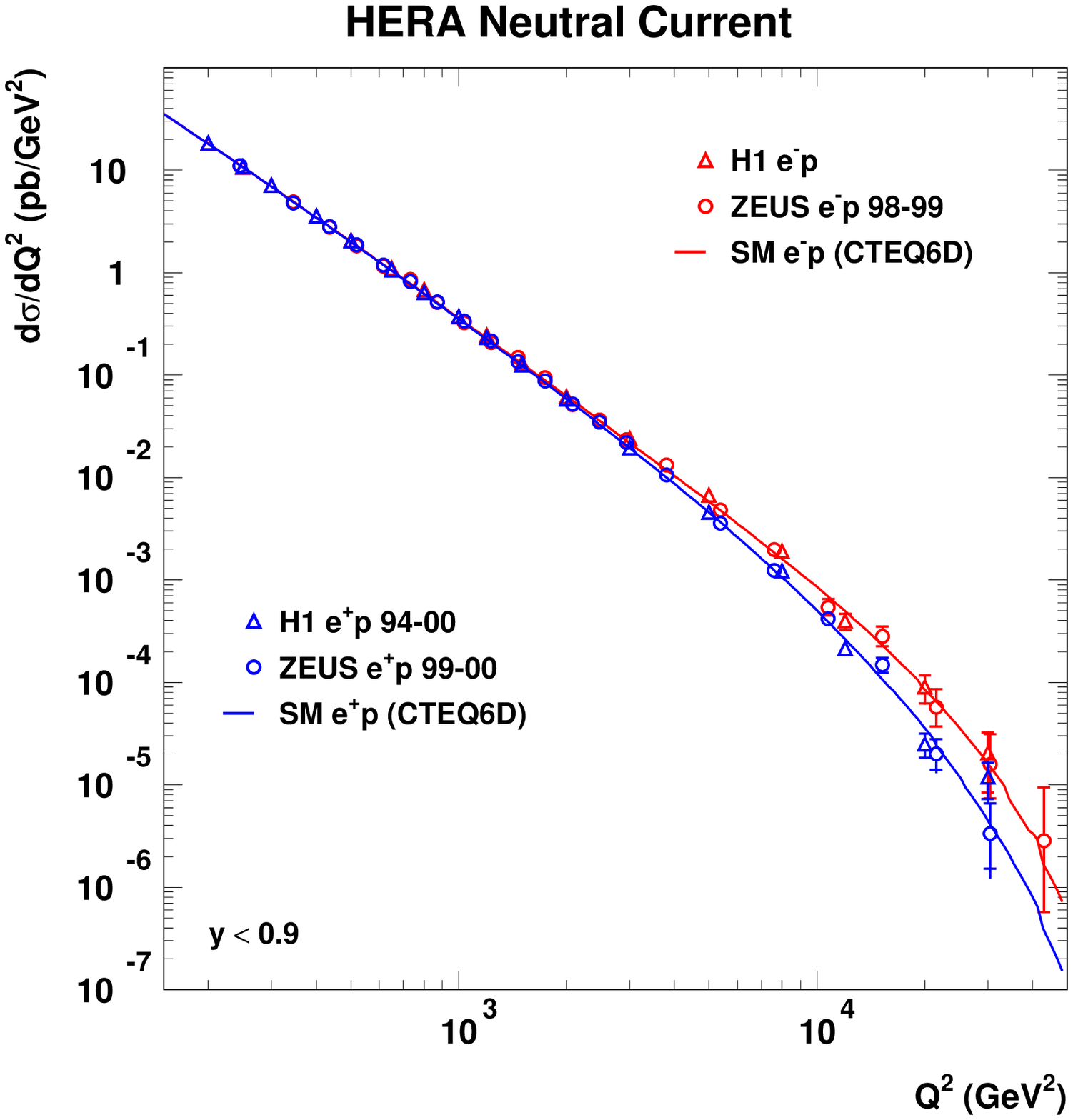,height=3.5in,width=3.0in}
\caption{The {NC} cross section versus $Q^2$ for $e^+p$ and $e^-p$ 
  interactions as measured by the {H1} and {ZEUS} experiments.}
\label{fig:NCepem}
\end{minipage}
\hspace*{0.5cm}
\begin{minipage}{8.0cm}
\psfig{figure=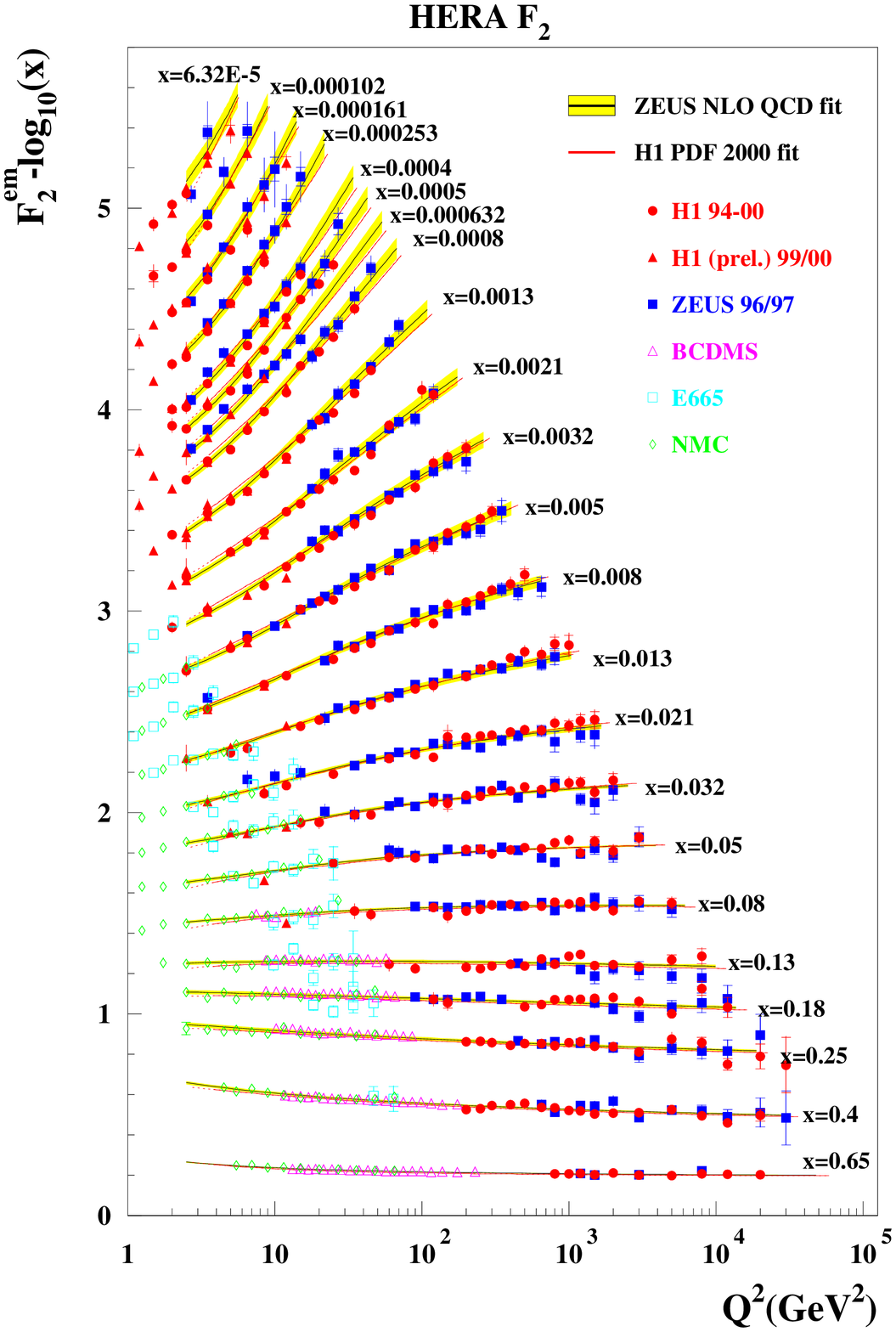,height=3.5in,width=3.0in}
\caption{The structure function $F_2^{em}$ versus $Q^2$ at
 fixed values of $x$ compared to a fit based on {DGLAP} evolution
 equations.}
\label{fig:NCf2}
\end{minipage}
\end{figure}

Fig.~\ref{fig:NCepem} , shows the {NC} cross section as a function of
$Q^2$ for $e^+p$ and $e^-p$ scattering as measured by the
{ZEUS}~\cite{ZEUS} and {H1}~\cite{H1} experiments. The {NC} cross
section for $e^+p$ is equal to that of $e^-p$ at low $Q^2$ because in
this kinematic region the electromagnetic interaction dominates. The
structure function $xF_3^{NC}$ has a negative contribution to the
cross section in case of $e^+p$ and a positive one in case of $e^-p$.
This can be observed in eq.~\ref{eq:dsnc}, where at high $Q^2$ the
difference between the $e^+p$ and $e^-p$ {NC} cross section becomes
noticeable.  Measurements by {H1} and {ZEUS} are in a good agreement
with expectations of the Standard Model ({SM}), calculated with the
{CTEQ6D}~\cite{Kretzer:2003it} parameterisation of parton distribution
functions ({PDFs}) in the proton.

The measurements of $F_2^{em}$ at {HERA I} are shown in
Fig.~\ref{fig:NCf2}, where the values of $F_2^{em}$ are plotted as a
function of $Q^2$ for fixed values of $x$. Strong scaling violations
are observed, especially at low values of $x$. This is the
manifestation of the presence of a large gluon density in the proton.
The observed scaling violations are well reproduced by the
{DGLAP}~\cite{dglap} evolution of perturbative {QCD}.

The structure function $xF_3^{NC}$ can be determined from the
difference of $e^{\pm}p$ cross sections,
\begin{eqnarray} 
  xF_3 = \frac{xQ^4}{4\pi \alpha^2 Y_-}\left 
    [\frac{d^2\sigma(e^-p)}{dxdQ^2} - 
    \frac{d^2\sigma(e^+p)}{dxdQ^2}\right ]\, .
\end{eqnarray}

\begin{figure}
\begin{center}
\psfig{figure=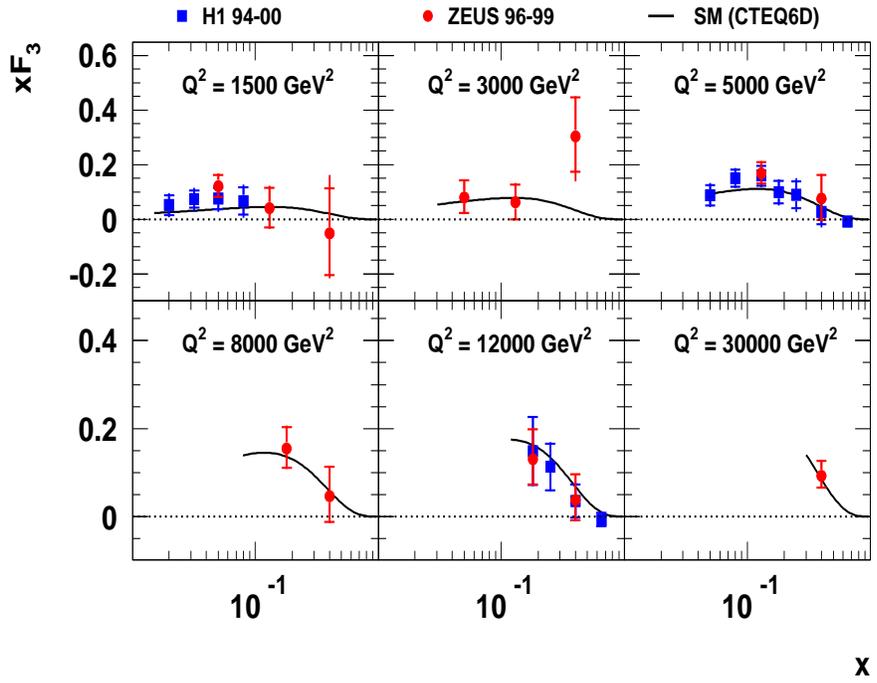,height=3.5in,width=4.5in}
\end{center}
\caption{The structure function $xF_3$ versus $x$ at different 
  values of $Q^2$ as obtained by the {H1} and {ZEUS} experiments.}
\label{fig:xf3}
\end{figure}

The $xF_3^{NC}$ extracted at {HERA I}~\cite{Chekanov:2002ej}$^,$
\cite{Adloff} is shown in Fig.~\ref{fig:xf3}, as a function of $x$ for
different $Q^2$ bins. Within uncertainties, the measured $xF_3^{NC}$
is in agreement with expectations obtained using the {CTEQ6D} {PDFs}.

\subsection{Charged current}

\begin{figure}
\begin{minipage}{8.0cm}
\psfig{figure=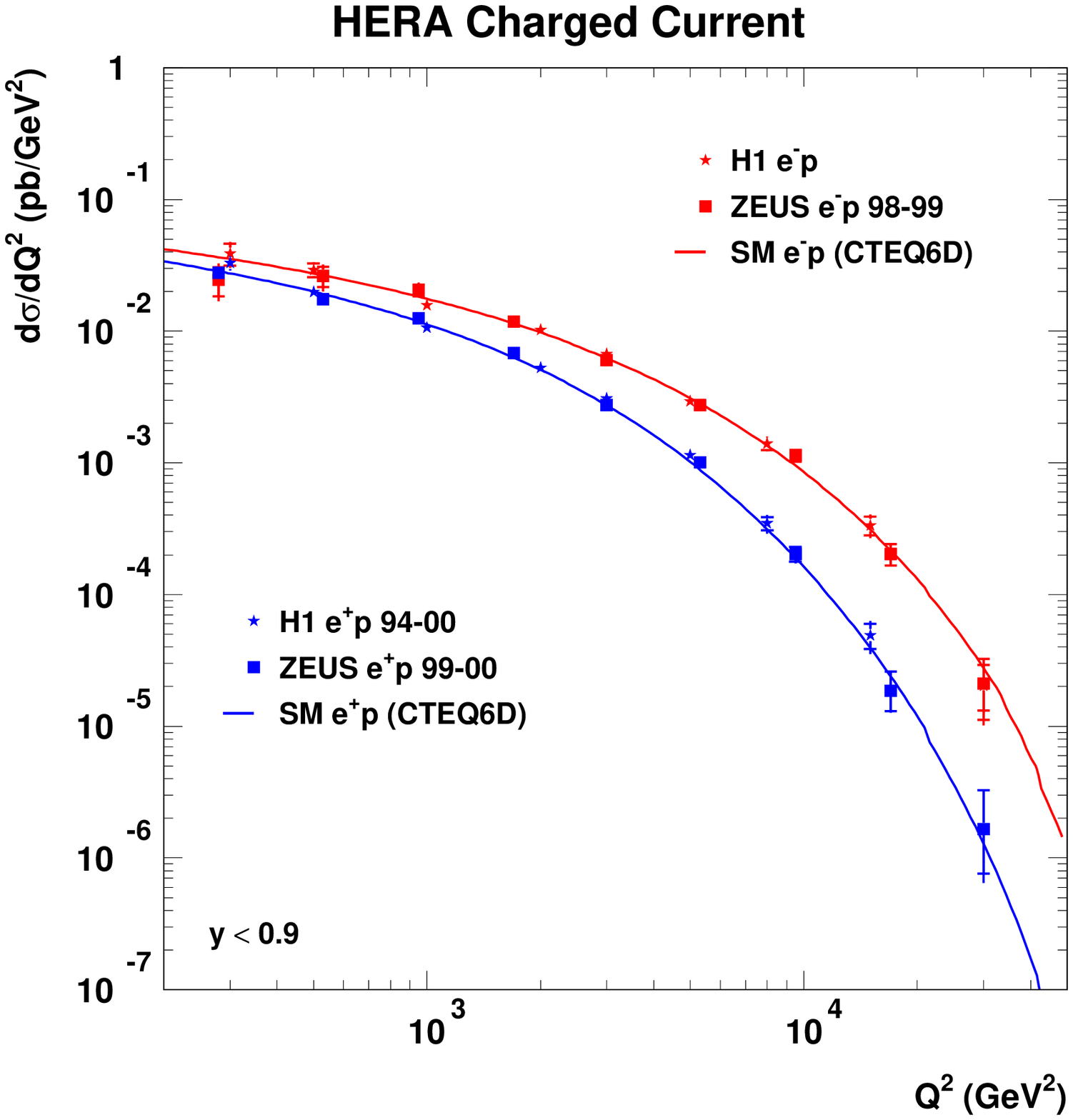,height=2.9in,width=3.in}
\caption{The {CC} cross section for $e^+p$ and $e^-p$ interactions 
  versus $Q^2$.}
\label{fig:CC_versus_Q2}
\end{minipage}
\hspace*{0.5cm}
\begin{minipage}{8.0cm}
\psfig{figure=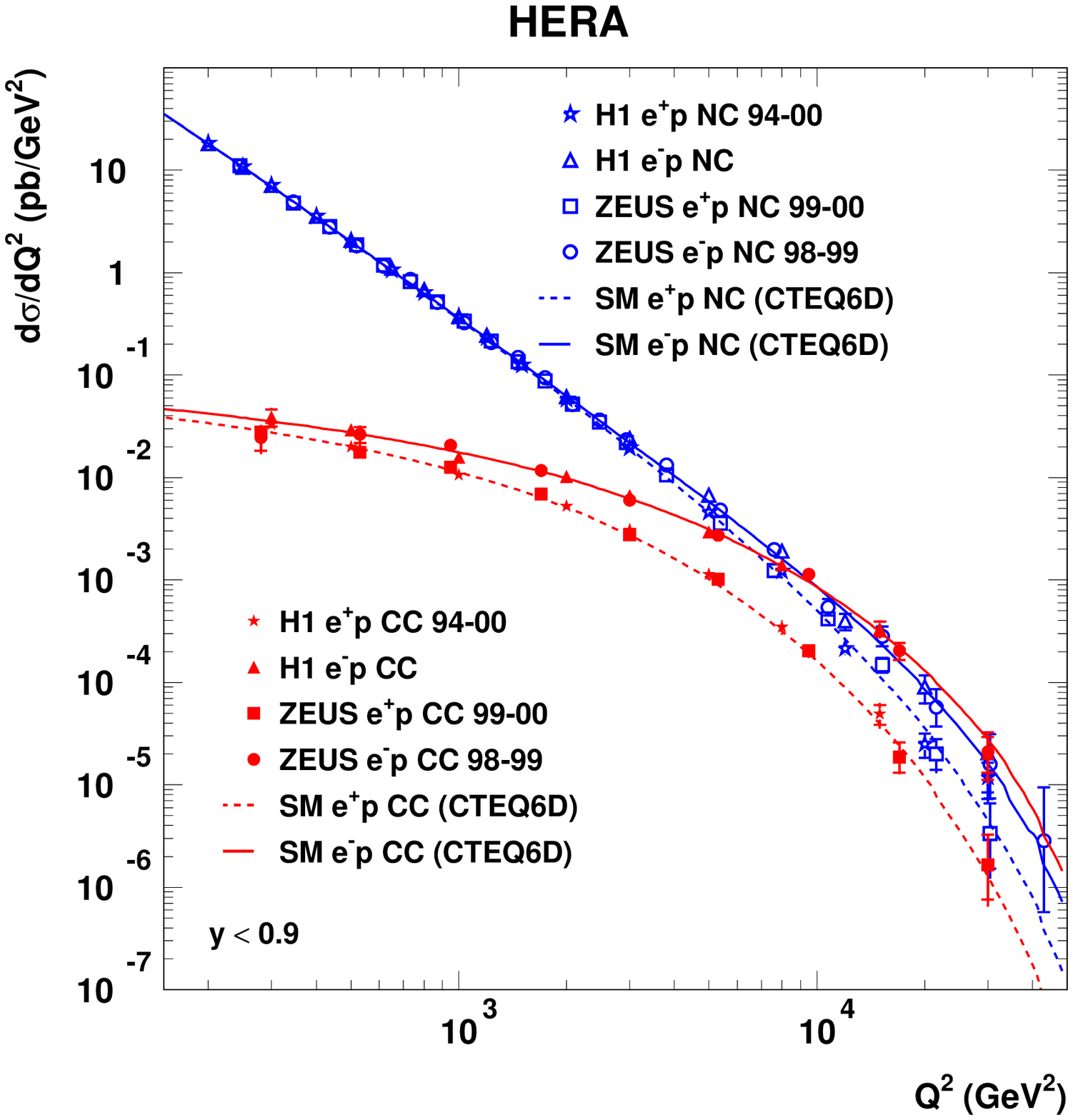,height=2.9in,width=3.in}
\caption{The {NC} and {CC} cross sections for
  $e^+p$ and $e^-p$ interactions versus $Q^2$. }
\label{fig:ew}
\end{minipage}
\end{figure}

The cross sections for {CC} deep inelastic scattering with
longitudinally unpolarised leptons on protons, $e^-p\rightarrow
\nu_eX$ and $e^+p\rightarrow \bar{\nu_e}X$, are given by
\begin{eqnarray}
  \frac{d^2\sigma_{Born}^{CC}(e^+p)}{dxdQ^2} = 
  \frac{G_{\mu}^2}{2\pi}\left (\frac{M_W^2}{Q^2+M_W^2}\right )^2\cdot 
       \left [(\bar{u}+\bar{c})+(1-y)^2(d+s)\right ] \, , 
\label{eq:dsccep}
\end{eqnarray}
\begin{eqnarray}
  \frac{d^2\sigma_{Born}^{CC}(e^-p)}{dxdQ^2} = 
  \frac{G_{\mu}^2}{2\pi}\left (\frac{M_W^2}{Q^2+M_W^2}\right )^2\cdot 
       \left [(u+c)+(1-y)^2(\bar{d}+\bar{s})\right ] \, , 
       \label{eq:dsccem}
\end{eqnarray}
where $G_{\mu}$ is the Fermi coupling constant and $M_W$ is the {W}
boson mass. The third quarks generation was neglected in
eq.~(\ref{eq:dsccep}) and in eq.~(\ref{eq:dsccem}). At high $x$
values, one can safely ignore the sea quarks and assume that the
proton is dominated by the valence quarks.  Under this assumption, the
$e^+p$ {CC} cross section is proportional to $(1-y)^2d$, while the
$e^-p$ {CC} cross section is proportional to $u$. Since the $u$ quark
dominates in the proton, the $e^-p$ {CC} cross section is higher than
the $e^+p$ {CC} cross section.  This is shown in
Fig.~\ref{fig:CC_versus_Q2}, where the $e^+p$ and $e^-p$ {CC} cross
section is plotted as a function of $Q^2$. A good agreement between
the two experiments, {H1} and {ZEUS}, is observed and the measured
values are well reproduced by the {SM} calculations based on the
{CTEQ6D} {PDFs}.

\subsection{EW unification}
A comparison of {NC} and {CC} $e^{\pm}p$ cross sections as a function
  of $Q^2$ is shown in Fig.~\ref{fig:ew}. Around
  $Q^2\sim 10^4~\mathrm{GeV^2}$, the {CC} and {NC} cross sections
  become equal. Since the {NC} cross section is dominated by
  electromagnetic component this equality can be intepreted as a
  manifestation of electroweak unification.

\section{HERA II}
\subsection{HERA II upgrade}
In order to study in detail the properties of deep inelastic
scattering at high $Q^2$, the {HERA} accelerator was upgraded to
deliver a five fold increase in luminosity and longitudinally
polarised lepton beams~\footnote{{HERA I} had provided longitudinal
  polarised lepton beam to the fixed target experiment, {HERMES}.}. To
meet the challenge, the {H1} and the {ZEUS} detectors were upgraded:
a new inner silicon vertex detector was installed at ZEUS to increase
the efficiency of the vertex reconstruction; an extension of the inner
silicon detectors in {H1} was done; new forward tracking detectors
were installed in both {H1} and {ZEUS} detectors; luminosity
detectors were modified in {H1} and {ZEUS}; a new track trigger
was added in {H1}~\cite{mehta}.

\begin{figure}
\begin{center}
\psfig{figure=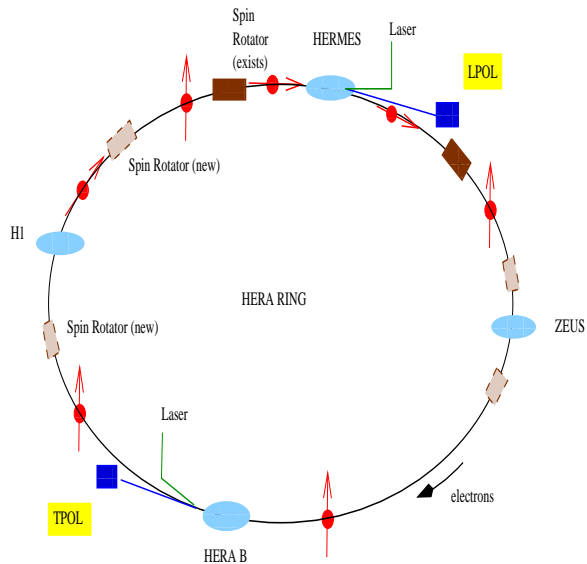,height=2.9in,width=3.in}
\end{center}
\caption{A schematic picture of {HERA II} after the upgrade. }
\label{fig:HERA}
\end{figure}

In order to measure the polarisation at {HERA II}, two independent
polarimeters are in use based on different measurement methods, the
longitudinal polarimeter ({LPOL}) and the transverse polarimeter
({TPOL}). The {LPOL} is located near the {HERMES} experiment and was
in use during the {HERA I} run period. The {TPOL} was upgraded and
is to be used in the {HERA II} run period. In addition, spin
rotators were installed so that both {H1} and {ZEUS} have now access
to longitudinally polarised $e^{\pm}$ beams. The location of the
polarimeters and the spin rotators is shown in Fig.~\ref{fig:HERA}.

\subsection{Charged current and polarisation}
The Standard Model predicts that the charged current cross
section for collisions of polarised electrons with protons is
proportional to $1\pm P$, where $P$ is the longitudinal polarisation
of electrons. In other words, the {SM} predicts that the {CC} cross
section for fully right-handed electrons (left-handed positrons) with
colliding protons is zero.

\begin{figure}
\begin{center}
\psfig{figure=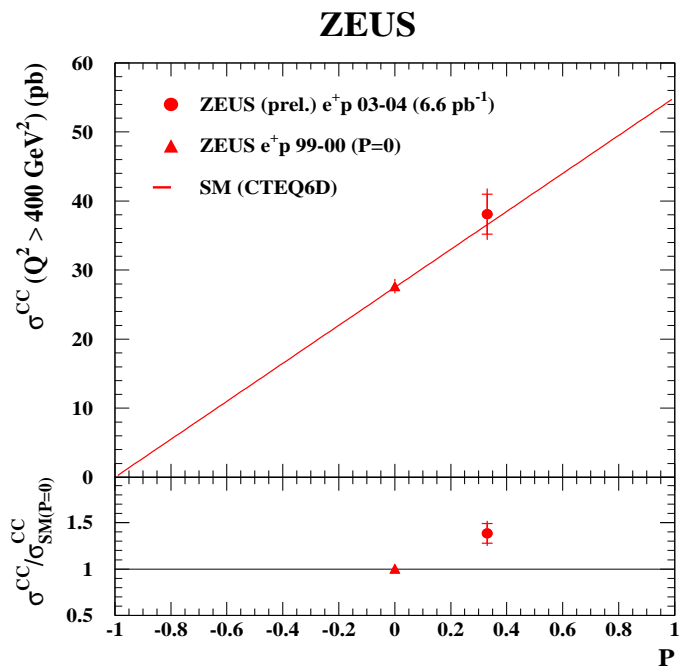,height=3.2in,width=3.5in}
\end{center}
\caption{Upper:the total {CC} {DIS} cross section for $e^+p$ 
        scattering plotted as a function of the polarisation of the
        incoming lepton beam, P.  Lower: the ratio of the polarised to
        unpolarised cross section as a function of the polarisation.}
\label{fig:pre_CC}
\end{figure}

The first measurement of $e^+p$ {CC} cross section with $P~=~33~\pm ~
2 \%$ and with integrated luminosity of $6.6~pb^{-1}$, has been made
by the {ZEUS} collaboration. The total {CC} cross section was measured
for $Q^2>400~\mathrm{GeV^2}$, and shows a significant rise of the
cross section relative to the unpolarised case from {HERA I}, as can
be seen in Fig.~\ref{fig:pre_CC}. In the lower part of
Fig.~\ref{fig:pre_CC}, the ratio of polarised to unpolarised cross
section is shown. The increase of the cross section is well reproduced
by the {SM} expectations.

\newpage

\section{Acknowledgements}
This work was partially supported by the Bj\"{o}rn Wiik World Laboratory
Scholarship Programme at DESY, the MINERVA Foundation, the Israel
Science Foundation (ISF) and the Israeli Ministry of Science.


\section{Bibliography}
\begin{thebibliography}{99}
\bibitem{HERA} HERA, ``A proposal for a large electron-proton
  colliding beam facility at DESY'', DESY HERA 81-10.
\bibitem{Klein} M. Klein and T. Riemann, Z. Phys. {\bf C 24}, 151
  (1984).
\bibitem{ZEUS} {ZEUS} Collab., S.~Chekanov {\it et al.}, ``High-$Q^2$
  neutral current cross sections in $e^+ p$ deep inelastic scattering
  at $\sqrt{s}~=~318$-GeV,''
  arXiv:hep-ex/0401003.\\
  {ZEUS} Collab., S.~Chekanov {\it et al.}, Eur.\ Phys.\ J.\ C
  {\bf 28}, 175 (2003).\\
  {ZEUS} Collab., S.~Chekanov {\it et al.},
  Eur.\ Phys.\ J.\ C {\bf 21}, 443 (2001).\\
  {ZEUS} Collab., J.~Breitweg {\it et al.}, Eur.\ Phys.\ J.\ C {\bf
    11}, 427 (1999).
\bibitem{H1} {H1} Collab., C.~Adloff {\it et al.}, Eur.\ Phys.\ J.\ C
  {\bf 13}, 609 (2000). \\
  {H1} Collab., C.~Adloff {\it et al.}, Eur.\ Phys.\ J.\ C {\bf 19},
  269 (2001). \\
  {H1} Collab., S.~Aid {\it et al.}, Phys.\ Lett.\ B {\bf 379}, 319
  (1996).
\bibitem{Kretzer:2003it} S.~Kretzer, H.~L.~Lai, F.~I.~Olness and
  W.~K.~Tung, JHEP {\bf 7}, 12 (2002).
\bibitem{dglap} V. N. Gribov and L. N. Lipatov, Sov. J. Nucl. Phys.
  {\bf 15}, 438 (1972). \\
  V. N. Gribov and L. N. Lipatov, Sov. J. Nucl. Phys. {\bf 15}, 675
  (1972).\\
  Yu.L. Dokshitzer, Sov. Phys. JETP {\bf 46}, 641 (1977).\\
  G. Altarelli and G. Parisi, Nucl. Phys. {\bf B 126}, 298 (1977).
\bibitem{Chekanov:2002ej} {ZEUS} Collab., S.~Chekanov {\it et al.},
  Eur.\ Phys.\ J.\ C {\bf 28}, 175 (2003)
\bibitem{Adloff} {H1} Collab., C. Adloff {\it et al.}, Eur.  Phys. J.
  {\bf C 19}, 269 (2001)
\bibitem{mehta} A. Mehta, HERA upgrade and prospects, ACTA Phys. Pol.
  {\bf B 33}, 3937 (2002).
\end{thebibliography}
\end{document}